\documentclass[12pt,preprint]{aastex}
\usepackage{epsfig}

\begin{document}

\newcommand{\gsim}{\hbox{\rlap{$^>$}$_\sim$}}
\voffset-.6cm

\title{Fireballs and cannonballs  confront the afterglow of  GRB
991208}

\author{Shlomo Dado\altaffilmark{1}, Arnon Dar\altaffilmark{1,2}
and A. De R\'ujula\altaffilmark{2}}

\altaffiltext{1}{Physics Department and Space Research Institute,
Technion, Haifa 32000, Israel} \altaffiltext{2}{Theory Division,
CERN,CH-1211 Geneva 23, Switzerland}


\begin{abstract}

Galama et al. have recently reported their follow-up measurements
of the radio afterglow (AG) of the Gamma Ray Burst (GRB) 991208,
up to 293 days after burst, and their reanalysis of the broad-band
AG, in the framework of standard fireball models. They advocate a
serious revision of their prior analysis and conclusions, based on
optical data and on their earlier observations during the first
two weeks of the AG.  We comment on their work and fill a lacuna:
these authors have overlooked the possibility of comparing their
new data to the available predictions of the cannonball (CB) model,
based ---like their incorrect predictions--- on the first round of
data.  The new data are in good agreement with these CB-model
predictions.  This is in spite of the fact that, in comparison to
the fireball models, the CB model is much simpler, much more
predictive, has many fewer parameters, practically no free choices...
and it describes well ---on a universal basis--- all the
measured AGs of GRBs of known redshift.

\end{abstract}


\section*{Introduction}

We discuss the afterglow of GRB 991208 as a good and simple example
of a three-sided confrontation:  the data, the generally accepted
theory (the fireball model in its various incarnations, thereafter
the ``standard model'': SM), and the cannonball (CB) model.

Gamma Ray Burst (GRB) 991208 was detected with the Interplanetary
Network (IPN) on December 8.192 UT, 1999 and its afterglow (AG)
was first detected on December 10.92 UT in the radio band (Hurley et
al. 2000).  The optical AG of GRB 991208 was first detected 2.1
days after burst and has been measured by Castro-Tirado et al.
(2001) and Sagar et al. (2000).  ``Early'' radio measurements, from
11.77 UT to 21.96 UT, December 1999 between 1.4 and 250 GHz were
reported by Galama et al. (2000) (thereafter G1) and were analyzed
in the framework of the SM along with the optical data.
Follow up observations, up to day $\rm 293$ after burst have been
recently reported and reanalyzed by Galama et al. (2002), (thereafter
G2), requiring a severe revision of their prior analysis and
conclusions.

We have analized the AG of GRB 991208 in the realm of the CB model
twice before: in Dado et al.  (2002a), thereafter DDD1, we fitted
the available R-band data on all GRBs of known redshift, $\rm z$,
including this one. In Dado et al. (2002b), hereinafter DDD2, we
extended the analysis to wide-band fits of all the available optical
and radio data.  We present here a comparison between these CB-model
predictions  and the new data of G2.  We also refit all the data
including the new observations in G2.  Consistently, the fit
parameters change
little as the new data are added.  More importantly, the CB model,
in spite of its simplicity and its scarcity of parameters and
choices, is found to be successful: the predictions of DDD2 agree
with the new data of G2, which is not the case for the SM
predictions in G1.

\section*{GRB 991208 secundum Galama et al. (2000) [G1]}

In G1 the authors first make a fit to the radio-to-optical spectral
behaviour of data modified to pertain to a fixed ``unified'' date:
December 1999, 15.5 UT. The ``spectral parameters'' are the
self-absorption and peak frequencies, $\rm \nu_a$ and $\rm \nu_m$,
the peak flux density, $\rm F_m$, and the power-law index of the
electron distribution, $\rm p$.  With the resulting $\rm p=2.52$
fixed, they fit the radio-only spectrum at 3 other unified
dates within the first 2 weeks, by extracting values of $\rm
\nu_a$, $\rm \nu_m$ and $\rm F_m$ separately for each of these
dates. This procedure is useful as a test of the time evolution
of these quantities but, {\it in totto}, theirs is a 13-parameter fit.

On the basis of their measured parameter-evolution, the authors of
G1 discard spherical-explosion models, either with a constant-density
interstellar medium (ISM), or a $\rm 1/r^2$ ``wind'' circumburst
profile (the ``ISM'' and ``WIND'' models).  They also note that a
``JET'' model disagrees by $>\! 4\sigma$ with the predicted power
decline of $\rm F_m$. They advocate a model with a transition from
a quasi-spherical to a jet evolution (Kumar and Panaitescu, 2000),
but they stop short of an explicit analysis on these grounds.  Yet,
they conclude, with no proof, that ``the jet model can account for
the [observations...] provided that the jet transition has not been
fully completed in the first two weeks after the event''.

The closing predictions in G1 are that $\rm \nu_a\propto t^{-14/13}$
at $\rm t\!>\! 10$ days, and the flux density $\rm F_\nu\propto
t^{-(2.2\; to \; 2.5)}$ for $\nu=8.46$ GHz at $\rm t\!>\! 12$ days,
as well as for $\nu=4.86$ GHz at $\rm t\!>\! 17$ days.

\section*{GRB 991208 secundum Galama et al. (2002) [G2]}

Two of the results of the analysis in G2 are that $\rm \nu_a\propto
t^{-0.29{-0.21\atop +0.17}}$ for a ``FREE fit'' to the ensemble of
data, and $\rm F_\nu\propto t^{-1.07\pm 0.09}$ at $\nu=8.46$ GHz,
for $\rm t=53$ to 293 days after burst. These results are in stark
contrast with the quoted predictions of G1.

In the FREE model of G2 one extra cooling frequency, $\rm \nu_c$,
is introduced, and  $\rm F_m$ as well as $\rm\nu_{a}$, $\rm \nu_m$
and $\rm \nu_c$ are assumed to behave as $\rm C_i\, t^{-\alpha_i}$,
for a total of nine parameters (including $\rm p$).  Three SM
variations are also discussed:  ISM, WIND and JET, each of which
predicts the values of $\rm \alpha_{_{F}}$, $\rm\nu_a$, $\rm \nu_m$
and $\rm\nu_c$.  The ISM and WIND models turn out to be inadequate,
as in G1.  The JET model (with its two new parameters) is an
improvement over that of G1, but fails to reconcile the late-time
decay $\rm F_\nu\sim t^{-1.1}$ at 8.46 GHz with the much steeper
optical decay $\rm F_R\propto t^{-2.2\pm 0.1}$ ---observed by Sagar
et al. (2000) at $\rm t=2$ to 10 days after burst--- which should
similarly decline. The FREE model provides a satisfactory fit to
the data, but implies that the combination $\rm\gamma\,B^3$ of the
bulk Lorentz factor of the flow and the post-shock magnetic field
ought to be roughly constant, while both are expected to decline
with time. In G2, the predictions by Li and Chevalier (2001) on
the late-time behaviour of this AG (in a model with two electron
energy distributions) are also found to fail.

Faced with so much unsuccess, the authors of G2 conclude that, in
analogy to work by Frail et al. (2000) on GRB 970508, ``the simplest
explanation which is consistent with the data and requires no
significant modifications is that the blast wave of GRB 991208
entered a non-relativistic expansion phase several months after
the burst''. As in G1, no explicit support is given to the conclusion.

\begin{figure}[t]
\begin{tabular}{cc}
\hskip 2truecm
\hspace*{-1.7cm}
\epsfig{file=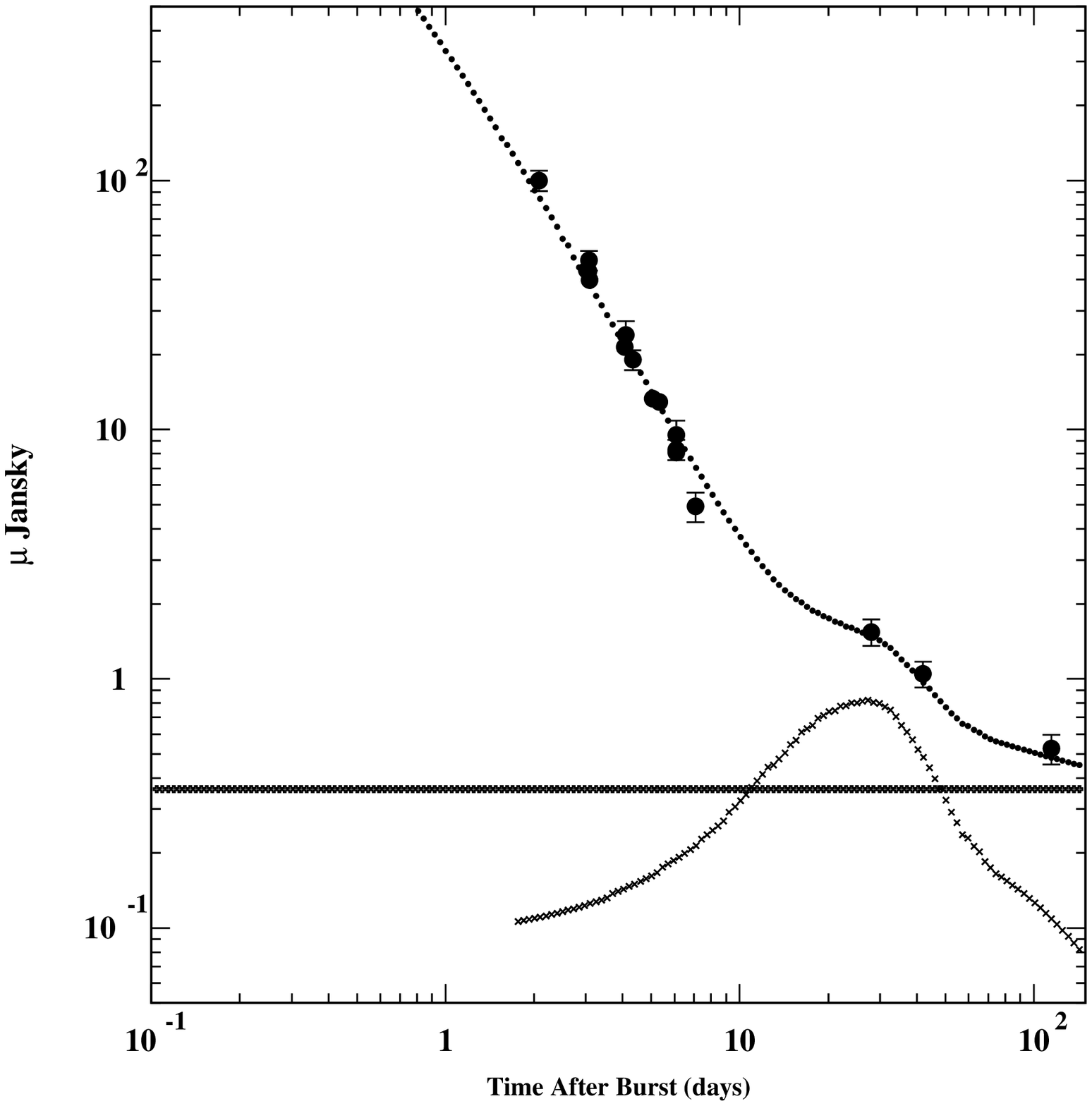, width=9cm} \\
\hspace*{.5cm}
\epsfig{file=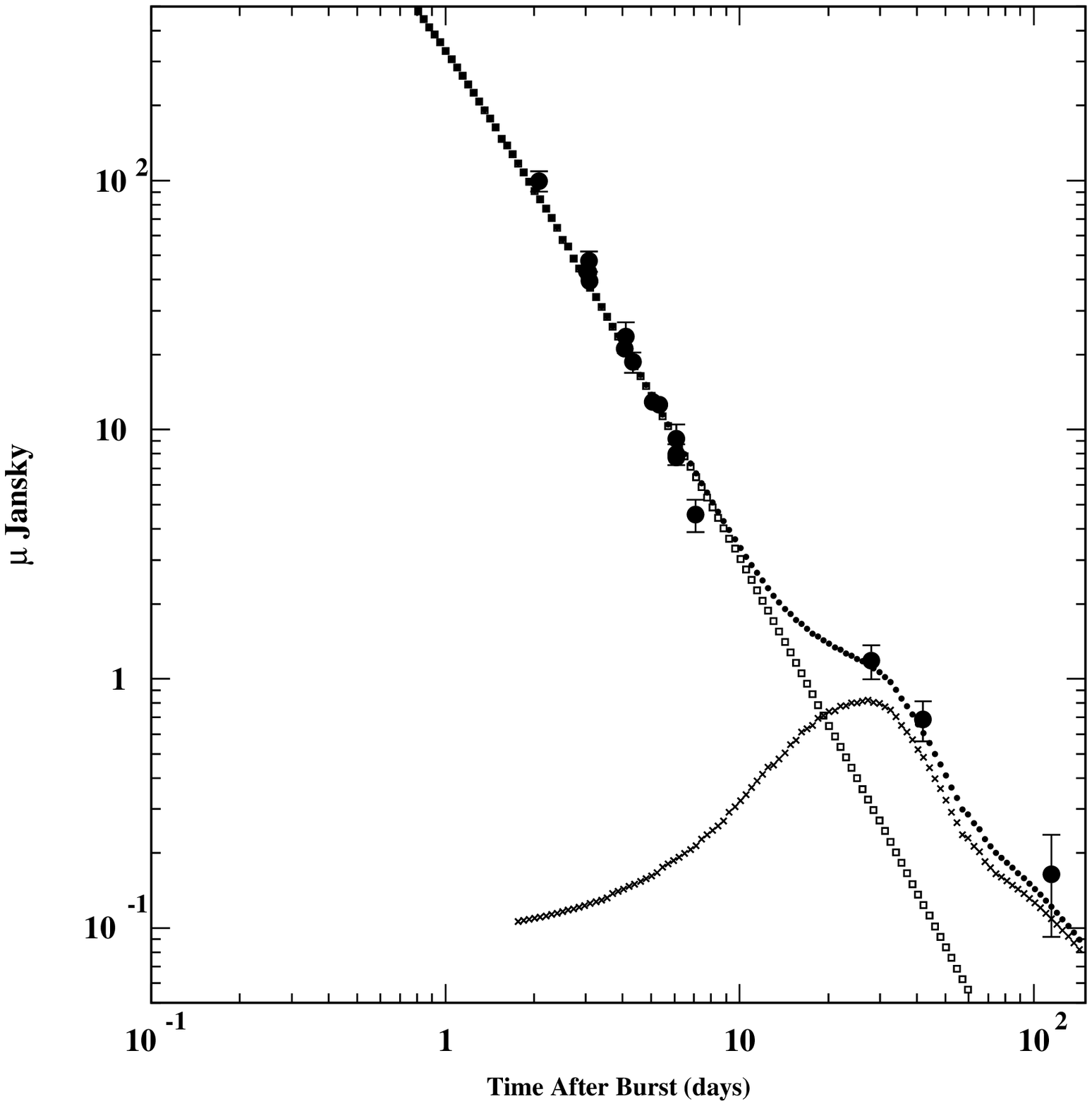, width=9cm}
\end{tabular}
\caption{Comparisons between the R-band AG
(upper curves) and the observations,
not corrected for  extinction,
for GRB 991208, at $\rm z=0.706$ (DDD1).
Upper panel: without subtraction of the host
galaxy's contribution (the straight line).
Lower panel: with the host galaxy subtracted. The contribution
from a 1998bw-like supernova placed at the GRB's
redshift, corrected for  extinction,
is indicated in both panels by a line of crosses.
The SN contribution is clearly discernible.}
\label{fone}
\end{figure}

The authors of G2 do not explain why they eliminate from their
analysis the optical measurements at $\rm t\!>\! 7$ days, a total
of 10 points in the R, V, B and I bands (Fig. 2 of Castro-Tirado
et al. 2001).  Just the two R-band points at days $\sim\! 24$ and
30 in our Fig. \ref{fone} are each $7 \sigma$ above the extrapolation
of the R-band fit in G2, after subtraction of the host galaxy
contribution ($\rm R=24.27\pm 0.13$, Castro-Tirado et al. 2001).
It is not obvious that the advocated late non-relativistic blast-wave
would remedy this discrepancy.

\section*{The parameters of the CB model}

In the CB model four parameters suffice to describe the {\it optical}
AGs in their various frequency bands. Three of them are ``intrinsic''
to the model: $\gamma_0$, the initial Lorentz factor of the CBs;
$\rm x_\infty$, the single parameter governing the deceleration of
a CB in the approximation of a constant-density interstellar medium
($\rm x_\infty/\gamma_0$ is the distance required to half the
original Lorentz factor); and an overall normalization. A fourth
parameter $\theta$, the angle between the line of sight to the
observer and the direction of the CBs, must be extracted from the
AG fits, but it is, in the same sense as the redshift, not a
parameter describing the model per se.

In extending the description of AGs from the optical to the radio
domain only one extra time-independent parameter is necessary: a
characteristic frequency for self-absorption within the CBs, $\rm
\nu_a$, for a total of 4 intrinsic parameters (DDD2). In DDD1 we
fit yet another parameter to the observations: the index $\rm p$
of the electron spectrum, prior to radiation losses. Having found
that, in the CB model, it was always compatible with the theoretical
expectation $\rm p\sim 2.2$, we no longer use it here, or elsewhere,
as a free parameter.

\section*{GRB 991208 in the CB model}

In Fig. \ref{fone} we show the fit of DDD1 to the R-band AG.  The
upper panel contains three contributions: the AG proper, the host
galaxy and a ``standard candle'' supernova akin to SN1998bw,
transported to the GRB's redshift\footnote{In the CB model, GRB
980425 ---associated to SN1998bw--- is in no way exceptional (Dar
and De R\'ujula 2000, DDD1, DDD2). Unlike in the SM, it makes sense
to use this SN as a putative standard candle.}.  In the lower panel
the galaxy is subtracted, demonstrating the presence of the SN: in
a CB-model analysis, in all instances wherein such a SN could be
seen, it was seen. This is so for all AGs with $\rm z\!<\!1.2$
(DDD1), including the cases where the presence of a 1998bw-like SN
was a prediction, based on the optical data {\it preceding} the
observable SN contribution (GRBs 011121 and 020405; Dado et al.
2002c,d).

\begin{figure}[t]
\begin{tabular}{cc}
\hskip 2truecm
\hspace*{-1.7cm}
\epsfig{file=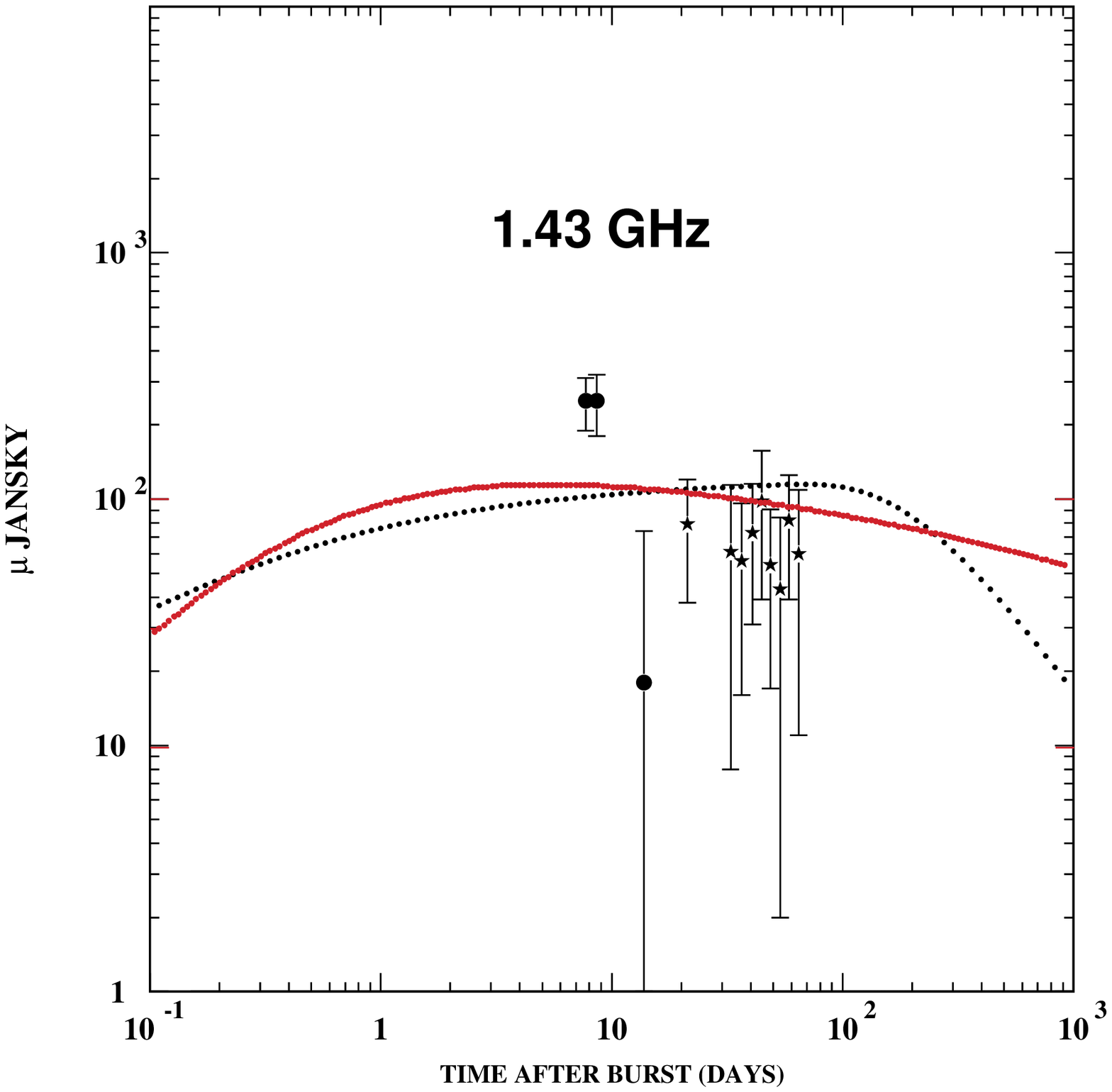, width=10cm} \\
\hspace*{.5cm}
\epsfig{file=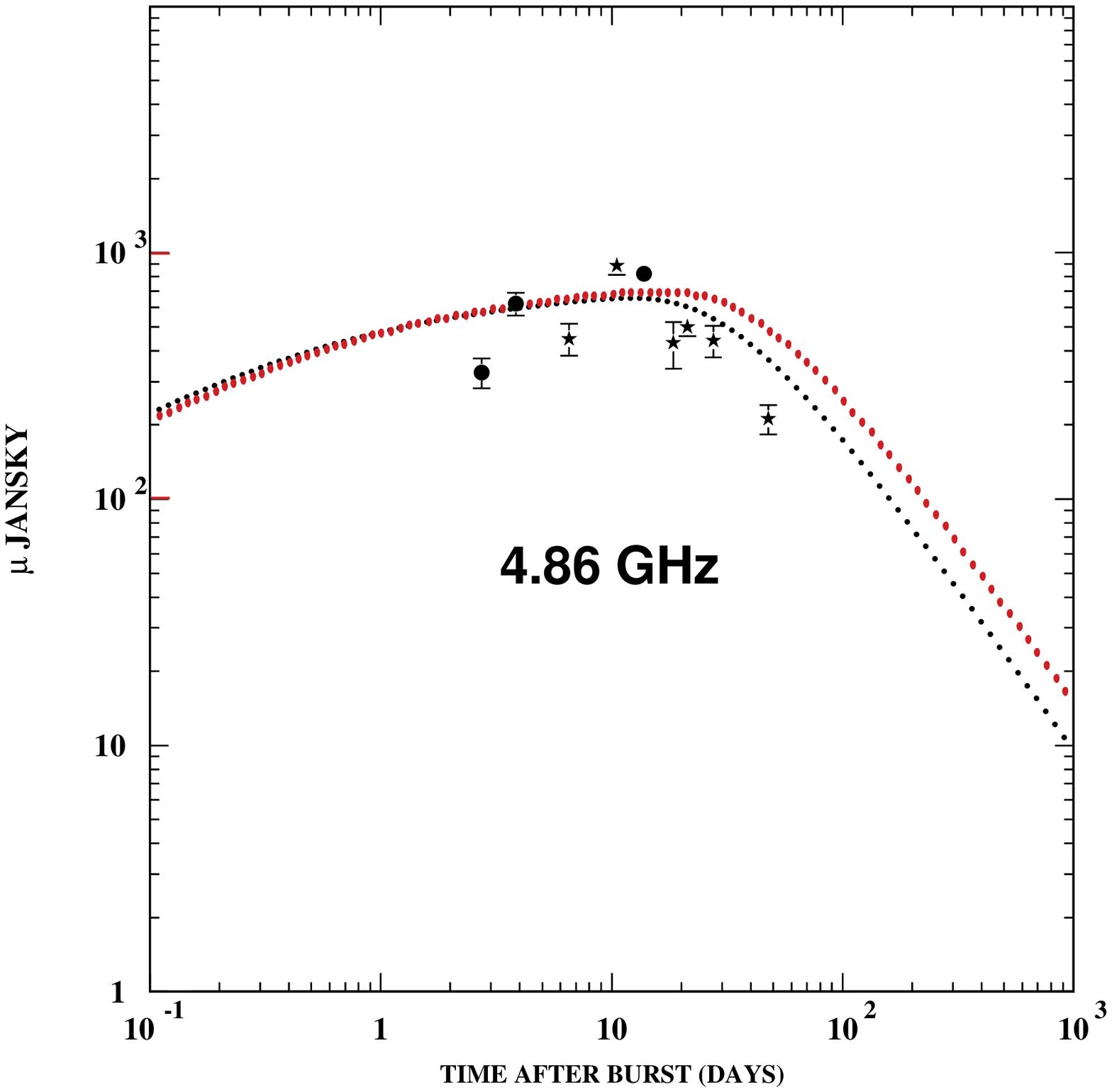, width=10cm}
\end{tabular}
\caption{Comparisons between the CB-model radio light-curves and
the observations at 1.43 and 4.86 GHz. The (red) continuous curve
in the upper panel and the (red) higher-up dotted line in the lower
panel are the results of DDD1, obtained without the new data,
represented by stars. The other dotted lines in both panels depict
the current fit to all data.}

\label{ftwo}
\end{figure}

\begin{figure}[t]
\begin{tabular}{cc}
\hskip 2truecm
\hspace*{-1.7cm}
\epsfig{file=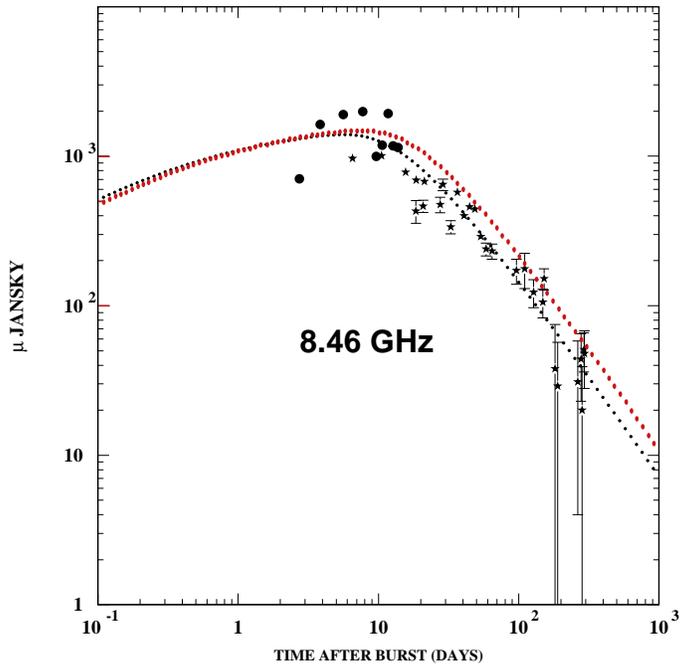, width=10cm} \\
\hspace*{.5cm}
\epsfig{file=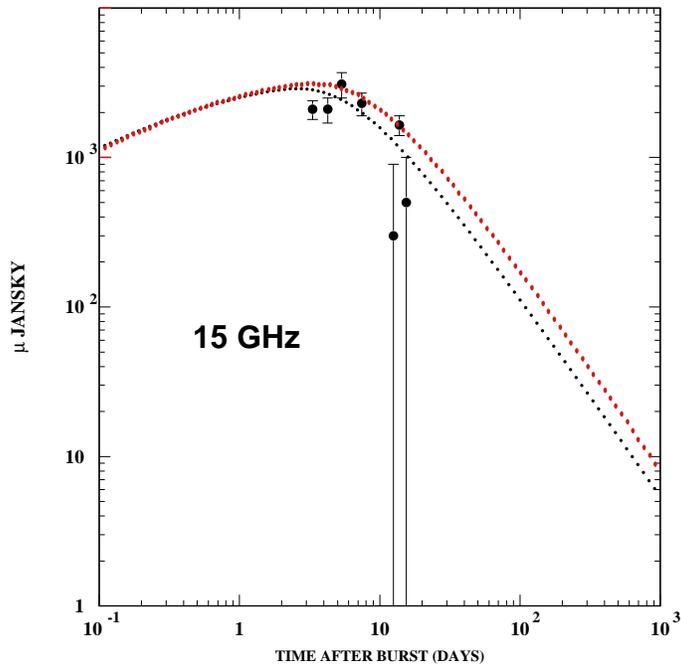, width=10cm}
\end{tabular}
\caption{Comparisons between the CB-model radio light-curves
and the observations at 8.46 and 15 GHz. The (red)
higher-up dotted lines are the results of DDD1, obtained
without the new data, represented by stars. The other dotted lines
in both panels depict the current fit to all data.}
\label{fthree}
\end{figure}

In Figs. \ref{ftwo} and \ref{fthree} we show two fits to the radio
data of GRB 991208, for $\nu=1.43$, 4.86, 8.46 GHz (for which there
is abundant new data in G2) and for 15 GHz, at which an earlier
measurement had escaped our attention in DDD2. One of these wide-band
fits (WB1) is the one published in DDD2, the other is a new fit
(WB2), along identical lines, including the new radio data of G2.
The figures show that the predictions of DDD2 were very satisfactory:
the WB1 and WB2 curves are very similar and they both provide a
good description of the data, their difference not being larger
than the scintillating ups and downs of the data.

In the Table we give the parameters for the R-band fit of DDD1,
the WB1 fit of DDD2 and the current WB2 fit. They appear to be
quite stable. Even the fit to only the R-band data determines
$\gamma_0$, $\theta$ and $\rm x_\infty$ to within a few percent of
the results of the WB2 fit (117 data points in total), even though
it is based on the mere dozen of early data points that are not
dominated by the SN. The value of $\rm x_\infty$ in the WB1 fit is
a bit smaller than in the others, the reason being that ---as can
be seen by inspection of Figs. \ref{ftwo}, \ref{fthree}--- this parameter is
sensitive to the late observations, and the early radio data of G1
dominated the WB1 fit. For fits that are so similar, their single
parameters describing the overall normalization are also necessarily
similar:  we have not reported them in the Table.

The CB model could be tested further by comparing the sky-projected
superluminal velocity of the CBs (that may be extracted from the AG's
radio scintillations, as for Galactic pulsars) with the predicted
$\rm v_{_{T}}(t)\simeq c\,\gamma(t)\,\delta(t)\, \theta/(1+z)$ (DDD2).

\section*{Asymptotic behaviours in the CB model}

According to G2 {\it ``one of the main challenges in modelling the AG
of GRB 991028''} is to reconcile the late radio decline
 at 8.46 GHz with the
optical decay, which is ``twice'' as fast. In the CB model this is not a
challenge, both behaviours are correctly predicted:

Let SEF refer to the ``proper'' CB wide-band
spectral energy flux (after subtraction
of the host galaxy and the associated SN).
Let $\rm\gamma=\gamma(t)$ be the explicit function describing the decreasing
Lorentz factor of the CBs (DDD1) and
$\delta\simeq 2\,\gamma/(1+\theta^2\gamma^2)$ the varying
Doppler factor of the radiation. The CB-model SEF has only two ``bends''.
 Self-absorption within the CBs results, in their rest system,
in an opacity $\rm\tau =(\nu_a/\nu)^2\,(\gamma/\gamma_0)^2$,
parametrized by the single parameter $\rm \nu_a$, and responsible
for the turn down of the SEF towards low $\nu$. At higher
$\nu$ the spectral index steepens from $\sim\! -1/2$ to $\rm -p/2$
at an ``injection bend'' frequency:
\begin{equation}
\rm \nu_b(t) \simeq \rm {1.87\times 10^{15}\over 1+z}\,
\left[{\gamma^3\, \delta\over 10^{12}}\right ]\,
\left[{n_p\over 10^{-3}cm^{-3}}\right]^{1/2}\;\; Hz,
\label{nubend}
\end{equation}
in the observer's frame, with $\rm n_p$ the ISM number density (DDD2).

After a couple of (observer) days, $\rm \gamma(t)\sim \delta(t)\sim t^{-1/3}$
and for frequencies above the opacity bend, or ``peak'', the SEF behaves
as:
\begin{eqnarray}
\rm F_{\nu(t)\ll \nu_b(t)}&\!\sim\!& \rm
\gamma^4\,
\nu^{-0.5}\sim
                t^{-1.33}\, \nu^{-0.5}\, , \\
\rm F_{\nu(t)\gg \nu_b(t)}&\!\sim\!& \rm
\gamma^{2p+2}\,\nu^{-p/2}\sim t^{-2.13}\,
             \nu^{-1.1}\, ,
\label{nuall}
\end{eqnarray}
where there may be a $\sim\!\pm 0.1 $ indetermination in the
the exponents of $\rm t$ and $\nu$,
due to the uncertainty around our adopted value, $\rm p=2.2$.
These predictions (or, rather, the explicit formula in DDD2 interpolating
them) are in agreement with the observations of all GRBs of known
$\rm z$. For the parameters that we fit to GRB 991208,
$\rm \nu(t)\!<\! \nu_b(t)$ in the radio at the late observed times,
and $\rm \nu(t)\!>\! \nu_b(t)$ in the optical, even during the
early optical observations.
In spite of their dependence on the arbitrarily chosen time intervals,
the observed late radio-behaviour at 8.46 GHz ($\rm \sim\! t^{-1.07\pm
0.09}$, G2) and the optical result ($\rm\sim\! t^{-2.3 \pm 0.07}\,
\nu^{-1.05\pm 0.09}$, Castro-Tirado et al. 2001; $\rm\sim\! t^{-2.2
\pm 0.1}$, Sagar et al. 2000) are compatible with the
above CB-model expectations, q.e.d.

\section*{Conclusions}

We have reviewed a particular example of the failure of the standard
model of GRBs in describing an afterglow (G1, G2), and the existence
of a much simpler, predictive and successful alternative. This is
not an exception, to date there is no satisfactory and comprehensive
SM explanation of the AGs of all GRBs of known redshift. In DDD1
we have commented on some of the most complete studies (Frail et
al. 2001; Kumar and Panaitescu 2001), which have, among others,
the limitation of being ``anthropoaxial'' (all jets point to the
observer, an unlikely circumstance).

Most researches in most areas of science are motivated by challenging
their respective ``standard models''. This is the case even for
models that are currently flawless, such as the SM of particle
physics, or even for ``sacred'' pillars of science, such as quantum
mechanics and general relativity. In studying G1 and G2, as well
as most of the current GRB literature, it is difficult to suppress
the impression that, for observers and theorists alike, the opposite
motivation prevails.  True enough, some of the profound inadequacies
of the ``fireball'' or ``firecone'' SMs models are occasionally
aired (e.g.  Ghisellini 2001, Lazzati 2002), but the final verdict
is always benevolent.  The fact that new epicycles must be added
with every ``novel'' observation is not unwelcome, even if the
additions are ponderous and totally ad-hoc, as is the case in the
interpretation of the $\rm Fe$ or other ``metal'' X-ray lines in
GRB AGs (reviewed in Lazzati 2002) or of the wiggly AG of GRB
021004 (Lazzati et al. 2002; Nakar et al. 2002; Heyl and Perna
2002).  In everyone of these cases the CB model offers an incredibly
simpler alternative (Dado et al. 2002e,f).

We are not saying that the CB model is entirely correct, it is a
simplification of what is no doubt a very complicated phenomenon;
it will either require modifications or  turn out to be completely
wrong. But its assumptions and predictions should be tested against
observations, or challenged for consistency. In other realms of
science the existence of a sensible model challenging the standard
lore would be very welcome, as opposed to ignored.

{\bf Acknowledgment:} This research was supported in part by the Helen
Asher Space Research Fund and by the VPR fund for research at the
Technion. One of us, Arnon Dar, is grateful for hospitality
at the CERN Theory Division.

{}

\newpage

{ \vskip 0.3 true cm
\noindent
{\bf Table:} Successive CB-model fits to the AG of GRB 991208.
R-band is a fit to only that optical frequency (DDD1). WB1 is the
wide-band fit in DDD2, with only the early radio-data. WB2 is
the current fit to all data.}
\vskip 0.5 true cm
\begin{table}[h]
\normalsize
\hspace{.0cm} 
\begin{tabular}{|l|c|c|c|c|}
\hline

Parameter & R-band & WB1 & WB2 \\
\hline
$\theta$ [mrad]           & 0.100 & 0.111    & 0.103  \\
$\gamma_0$            & 1034  &  1034    & 1089   \\
$\rm x_{\infty} $ [Mpc]            & 1.357 & 1.014    & 1.382  \\
$\rm\nu_a $ [MHz]           & ****      &   103    &    89  \\
\hline

\end{tabular}
\end{table}

\end{document}